\documentclass[letterpaper]{jpconf}
\usepackage{graphicx}
\usepackage{amssymb}
\begin{document}
\title{Hans Bethe: The Nuclear Many Body Problem}

\author{Jeremy W. Holt and Gerald Brown}

\address{Department of Physics and Astronomy, State University of New York,
  Stony Brook, NY 11794-3800, USA}


\section{The Atomic and Nuclear Shell Models}
Before the second world war, the inner workings of the nucleus were a mystery.
Fermi and collaborators in Rome had bombarded nuclei by neutrons, and the
result was a large number of resonance states, evidenced by sharp peaks in the
cross section; i.e., in the off-coming neutrons. These peaks were the size of
electron volts (eV) in width. (The characteristic energy of a single molecule
flying around in the air at room temperature is the order of 1/40 of an 
electron volt. Thus, one electron volt is the energy
of a small assemblage of these particles.) On the other hand, the difference
in energy between low-lying energy levels in light and medium nuclei is the
order of MeV, one million electron volts.

Now Niels Bohr's point \cite{bohr1} was that if the widths of the nuclear
levels (compound states) were more than a million times smaller than the
typical excitation energies of nuclei, then this meant that the neutron
did not just fall into the nucleus and come out again, but that it collided
with the many particles in the nucleus, sharing its energy. According to the 
energy-time uncertainty relation, the width $\Delta E$ of such a
resonance is related to the lifetime of the state by \mbox{$t\simeq
  \hbar/\Delta E\sim \hbar/(1$  eV)} . In fact,
the time \mbox{$t\sim \hbar /(10$ MeV)} is the characteristic time for a
nucleon to 
circle around once in the nucleus, the dimension of which is Fermis (\mbox{1
  Fermi} = \mbox{$10^{-13}$ cm}). Thus, a thick ``porridge'' of
all of the nucleons in 
the nucleus was formed, and only after a relatively long time would this
mixture come back to the state in which one single nucleon again possessed all
of the extra energy, enough for it to escape. The compound states in
which the incoming energy was shared by all other particles in the porridge
looked dauntingly complicated.

Now atomic physics was considered to be very different from nuclear physics,
the atomic many body problem being that of explaining the makeup of
atoms. Niels 
Bohr \cite{bohr2} had shown, starting with the hydrogen atom, that each
negatively charged electron ran around the much smaller nucleus in one of many
``stationary states.'' Once in such a state, it was somehow ``protected'' from
spiraling into the positively charged nucleus. (This ``protection'' was
understood only later with the discovery of wave mechanics by Schr\"odinger
and Heisenberg.) The allowed states of the hydrogen atom were obtained in the
following way. The Coulomb attraction between the electron and proton
provides the centripetal force, yielding
\begin{equation}
\frac{e^2}{r^2} = \frac{mv^2}{r},
\label{centrip}
\end{equation}
where $-e$ is the electron charge, $+e$ is the proton charge, $m$ is the
electron 
mass, $v$ is its velocity, and $r$ is the radius of the orbit. Niels Bohr
carried out the quantization, the meaning of which will be clear later, using
what is called the classical action, but a more transparent (and equivalent)
way is to use the particle-wave duality picture of de Broglie (Prince
\mbox{L.\ V.\ de Broglie} received the Nobel prize in 1929 for his discovery
of the wave nature of electrons.) in which a particle with mass $m$ and
velocity $v$ is assigned a wavelength
\begin{equation}
\lambda = \frac{h}{mv},
\end{equation}
where $h$ is Planck's constant, ubiquitous in quantum mechanics. If
the wave is to be stationary, it must fit an integral number of times around
the circumference of the orbit, leading to
\begin{equation}
n\frac{h}{mv} = 2 \pi r.
\label{circum}
\end{equation}
Eliminating $v$ from eqs.\ (\ref{centrip}) and (\ref{circum}), we find the
radius of the orbit to be 
\begin{equation}
r = \frac{n^2 \hbar^2}{me^2}, \mbox{ \hskip.2in where \hskip.2in} \hbar =
\frac{h}{2\pi}.
\label{radius}
\end{equation}
Now the kinetic energy of the electron is
\begin{equation}
T = \frac{1}{2}mv^2 = \frac{e^2}{2r},
\label{kinetic}
\end{equation}
where we have used eq.\ (\ref{centrip}). The potential energy is
\begin{equation}
V = -\frac{e^2}{r} = -2T,
\label{poten}
\end{equation}
which follows easily from \mbox{eq.\ (\ref{kinetic})}. \mbox{Eq.\
  (\ref{poten})} is known as 
a ``virial relation,'' an equation that relates the kinetic and potential
energies to one another. In the case of a potential depending on $r$ as 
$1/r$, the kinetic energy is always equal to $-1/2$ of the potential
energy. From this relation we find that the total energy is given by
\begin{equation}
E = T + V = -T.
\label{etotal}
\end{equation}
Finally, one finds by combining \mbox{eqs.\ (\ref{etotal})}, (\ref{kinetic}),
and (\ref{radius}) that 
\begin{equation}
E_n = -\frac{1}{2}\frac{me^4}{n^2\hbar^2} = -\frac{1}{n^2}Ry,
\end{equation}
where $Ry$ is the Rydberg unit for energy
\begin{equation}
Ry \simeq 13.6 \mbox{ electron volts.}
\end{equation}
Thus, we find only a discrete set of allowed energies for an electron bound in
a hydrogen atom, and we label the corresponding states by their values of
$n$. Since \mbox{$n = 1$} for the innermost bound orbit in hydrogen, called an
$s$-state for reasons we discuss later, 13.6 electron volts is the ionization
energy, the energy necessary to remove the electron.

Electrons are fermions, such that only one particle can occupy a given quantum
state at a time. This is called the ``exclusion principle.'' (Wolfgang Pauli
received the Nobel prize in 1945 for the discovery of the exclusion principle,
also called the ``Pauli principle.'') Once an electron has been put into a
state, it excludes other electrons from occupying it. However, electrons have
an additional property called spin, which has the value 1/2 (in units of
$\hbar$), 
and this spin can be quantized along an arbitrary axis to be either up or
down. Thus two electrons can occupy the $1s$ state. Putting two electrons
around a nucleus consisting of two neutrons and two protons makes the helium
atom. Helium is the lightest element of the noble gases. (They are called
``noble'' because of their little chemical interaction.) Since the $1s$ shell
is filled with two electrons, the helium atom is compact and does not have an
empty $1s$ orbital which would like to grab a passing electron, unlike
hydrogen which is very chemically active due to a vacancy in its $1s$
shell.

To go further in the periodic table we have to put electrons into the \mbox{$n
  = 2$} orbit. A new addition is that an electron in the \mbox{$n = 2$} state
  can have an 
orbital angular momentum of \mbox{$l = 0$} or $1$, corresponding to the 
states $2s$ and $2p$, respectively. (The spectroscopic ordering is
$s\,p\,d\,f\,g\,h\,i$ 
for $l=0-6$, a notation that followed from the classification of atomic spectra
well before the Bohr atom was formulated.) The angular momentum 
\mbox{$l = 1$} of the $2p$ state can be projected on an arbitrary axis to give
components \mbox{$m = 1,0,$} or $-1$. So altogether, including spin, six
  particles can 
be put into the $2p$ state and two into the $2s$ state. Thus, adding eight
electrons in the $2s$- and $2p$-states, the next member of the noble gases,
neon, is obtained. It is particularly compact and the electrons are well
bound, because both the \mbox{$n = 1$} and \mbox{$n=2$} shells are filled.

Consider an element in which the $n=2$ shell is not filled, oxygen, which
has eight electrons, two in the \mbox{$n=1$} shell and six in the \mbox{$n=2$}
shell. 
Oxygen in the bloodstream or in the cellular mitochondria is always on
the alert to fill in the two empty orbits. We call such a ``grabbing''
behavior ``oxidation,'' even though the grabbing of electrons from other
chemical elements is done not only by oxygen. The molecules that damage living
cells by stealing their electrons are called ``free radicals,'' and it is
believed that left alone, they are a major cause of cancer and other
illnesses. This is the origin of the term ``oxidative stress.'' We
pay immense amounts of money for vitamins and other ``antioxidants'' in order
to combat free radicals by filling in the empty states.

We need to add one further piece to the picture of the atomic shell model, the
so-called $j-j$ coupling, which really gave the key success of the nuclear
shell model, as we explain later. The $j$ we talk about is the total angular
momentum, composed of adding the orbital angular momentum $l$ and the spin
angular momentum $s$. The 
latter can take on projections of $+1/2$ or $-1/2$ along an arbitrary
axis, so $j$ can be either $l+1/2$ or $l-1/2$. The possible projections of
$j$ are $m=j,j-1,\dots,-j$. Thus, if we reconsider the $p$-shell in an atom,
which has $l=1$, the projections are reclassified to be those of $j=3/2$ and
$j=1/2$. The former has projections $3/2,1/2,-1/2,$ and $-3/2$, the
projections differing by integers, whereas the latter has $+1/2$ and
$-1/2$. Altogether there are six states, the same number that we found
earlier, through projections of $m_l=-1,0,1$ and of $m_s=+1/2,-1/2$. The
classification in terms of $j$ is important because there is a spin-orbit
coupling; i.e., an interaction between spin and orbital motion which depends
on the relative angle between the spin and orbital angular momentum. This
interaction makes it more favorable for the spin to be aligned opposite to the
direction of the orbital angular momentum, as in the $2p_{3/2}$ state being
higher in energy than the $2p_{1/2}$ state. In this notation, the subscript
refers to the total angular momentum. The filling of the shells in the
$j-j$ classification scheme is the same as in the $l$ scheme in
that the same number of electrons fill a shell in either scheme, the subshell
labels being different, however.

The above is the atomic shell model, so called because electrons are filled in
shells. The electric interaction is weak compared with the nuclear one, down 
by the factor of $\alpha = e^2/\hbar c = 1/137$ from the nuclear interaction
(called the strong interaction). The atomic shell model is also determined
straightforwardly because the nucleus is very small (roughly 1,000 times
smaller in radius than the atom), and it chiefly acts as a
center about which the electrons revolve. In atoms the number of negatively
charged electrons is equal to the number of 
positively charged protons, the neutrons being of neutral charge. In light
nuclei there are equal numbers of neutrons and protons, but the number of
neutrons relative to the protons grows as the mass number
$A$ increases, because it is relatively costly in energy to concentrate the
repulsion of the protons in the nucleus. On the whole, electric forces play
only a minor role in the forces between nucleons inside the nucleus, except
for determining the ratio of protons to neutrons. By the time we get to
$^{208}$Pb with 208 nucleons, 126 neutrons and 82 protons, we come to a
critical situation for the nuclear shell model, which requires the spin-orbit
force, as we discuss later.

After a decade or two of nuclear ``porridge,'' imagine people's surprise when
the nuclear shell model was introduced in the late 1940's \cite{may,jen} and
it worked; i.e., it explained a lot of known nuclear characteristics,
especially the ``magic numbers.'' That is, as shells were filled in a
prescribed 
order, those nuclei with complete shells turned out to be substantially
more bound than those in which the shells were not filled. In particular, a
particle added to a closed shell had an abnormally low binding energy.

What determines the center in the nuclear shell model? In the case of the
atomic shell model, the much heavier and much smaller nucleus gave a center
about which the electrons could be put into shells. Nuclei were known to be
tightly bound. There must be a center, and most simply 
the center would be exactly in the middle of the charge distribution. This
charge distribution can be determined by considering two nuclei that differ
from one another by the exchange of a proton and a neutron. In this case
the two nuclei will have different binding energies due to the extra
Coulomb energy associated with the additional proton. This difference could be
measured by the  
energy of the radioactive decay of the one nucleus into the other and
estimated roughly as $Ze^2/R$ (where $R$ is the radius of the nucleus and $Z$
is the number of protons). This
gave $R \sim 1.5A^{1/3}\times 10^{-13}$ cm, where $A$ is the mass
number. (Later electron scattering experiments, acting
like a very high resolution electron scattering microscope, gave the detailed
shape of the charge distribution and basically replaced the 1.5 by 1.2.) Since
we now have an estimate of the nuclear charge distribution, we would like to
use this information to help determine the shell model potential.

Now the most common force is zero force, that is, matter staying at rest in
equilibrium. The force, according to Newton's law, is the (negative)
derivative of the potential. If the potential at short distances is some
constant times the square of $r$; i.e., $V=Cr^2$, then the force
$F=-dV/dr=-2Cr$ is zero at $r=0$. Furthermore, the negative sign indicates
that any movement away from $r=0$ is met by an attractive force directed back
toward the center, leading to stable equilibrium. Such a potential is called a
harmonic oscillator. It occurs 
quite commonly in nature, since most matter is more or less in equilibrium.

\begin{figure}[ht]
\begin{center}
\begin{minipage}{17pc}
\begin{center}
\includegraphics[width=10pc]{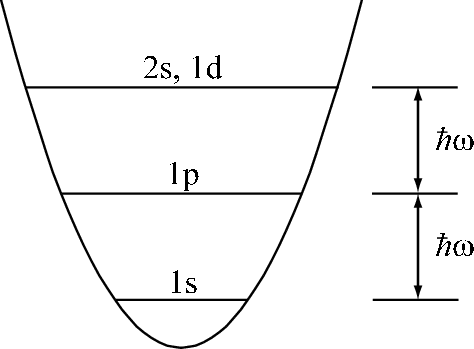}
\end{center}
\caption{Harmonic oscillator potential with possible single-particle states.}
\label{harmoscil1}
\end{minipage}\hspace{2pc}
\begin{minipage}{17pc}
\begin{center}
\includegraphics[width=10pc]{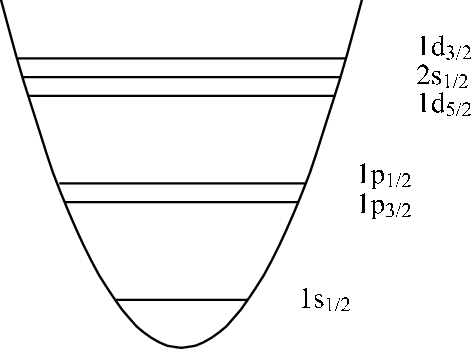}
\end{center}
\caption{Harmonic oscillator potential with possible single particle states in
 $j$--$j$ coupling.} 
\label{harmoscil2}
\end{minipage}
\end{center}
\end{figure}

The energy of a particle in such a potential can be expressed classically as
\begin{equation}
E=\frac{p^2}{2M}+\frac{1}{2}M\omega^2r^2,
\end{equation}
where $p$ is its momentum, $M$ is its mass, and $\omega$ is a parameter that
roughly describes the width of the potential well. Then one can define a
length parameter $a$ through  
\begin{equation}
\hbar \omega = \frac{\hbar^2}{Ma^2}
\end{equation}
and the root mean square radius $R$ of the nucleus can be determined in terms
of $A$ and $a$, $R$ being proportional to $A^{1/3}$. Using harmonic oscillator
wave functions with $a$ (or $\omega$) determined so that
\begin{equation}
R=1.2A^{1/3}\times 10^{-13} \mbox{ cm}
\end{equation}
gives a remarkably good fit to the shape of the charge distribution (in which
only the protons are included) of the nucleus. The possible energy levels in
such a potential are shown in \mbox{Fig.\ \ref{harmoscil1}} where we have
labeled them 
in the same way as in the atomic shell model. Note that the distances between
neighboring levels are always the same, $\hbar \omega$.\footnote{The
  situation 
  is analogous to that with light quanta, which each have energy $\hbar
  \omega$, so that an integral number of quanta are emitted for any given
  frequency $\hbar \omega$.}
We redraw the potential in \mbox{Fig.\ \ref{harmoscil2}}, incorporating the
$j-j$ coupling. 

Note that the spin-orbit splitting, that is, the splitting between two states
with the same $l$, is somewhat smaller than the
distance between shells\footnote{In heavier nuclei such as $^{208}$Pb
  the angular momenta become so large that the spin-orbit splitting is
  not small compared to the shell spacing, as we shall see.},
so that the classification of levels according to $l$ as shown in
Fig.\ \ref{harmoscil1} gives a good zero-order description.

At first sight the harmonic oscillator potential appears unreasonable because
the force drawing the particle back to the center increases as the particle
moves farther away, but nucleons can certainly escape from nuclei when given
enough energy. However, the
last nucleon in a nucleus is typically bound by about 8 MeV, and this has the
effect that its wave function drops off rapidly outside of the potential;
i.e., the probability of finding it outside the potential is generally
small. In fact, the harmonic oscillator potential gives remarkably good wave
functions, which can only be improved upon by very detailed calculations.

In the paper of the ``Bethe Bible'' coauthored with R.\ F.\ Bacher
\cite{bethbach}, nuclear masses had been measured accurately enough in the
vicinity of $^{16}$O so that ``It may thus be said safely that the completion
of the
neutron-proton shell at $^{16}$O is established beyond doubt from the data
about nuclear masses.'' The telltale signal of the closure of a shell is that
the binding energy of the next particle added to the closed shell nucleus is
anomalously small. Bethe and Bacher (p.\ 173) give the shell model levels in
the harmonic 
oscillator potential, the infinite square well, and the finite square
well. These figures would work fine for the closed shell nuclei $^{16}$O and
$^{40}$Ca. In Fig.\ \ref{harmoscil1}, $^{16}$O would result from filling the
$1s$ and $1p$ shells, $^{40}$Ca from filling additionally the $2s$ and $1d$
shells. However, other nuclei that were known to be tightly bound, such as
$^{208}$Pb, could not be explained by these simple potentials.

\begin{figure}
\begin{center}
\includegraphics[height=6in]{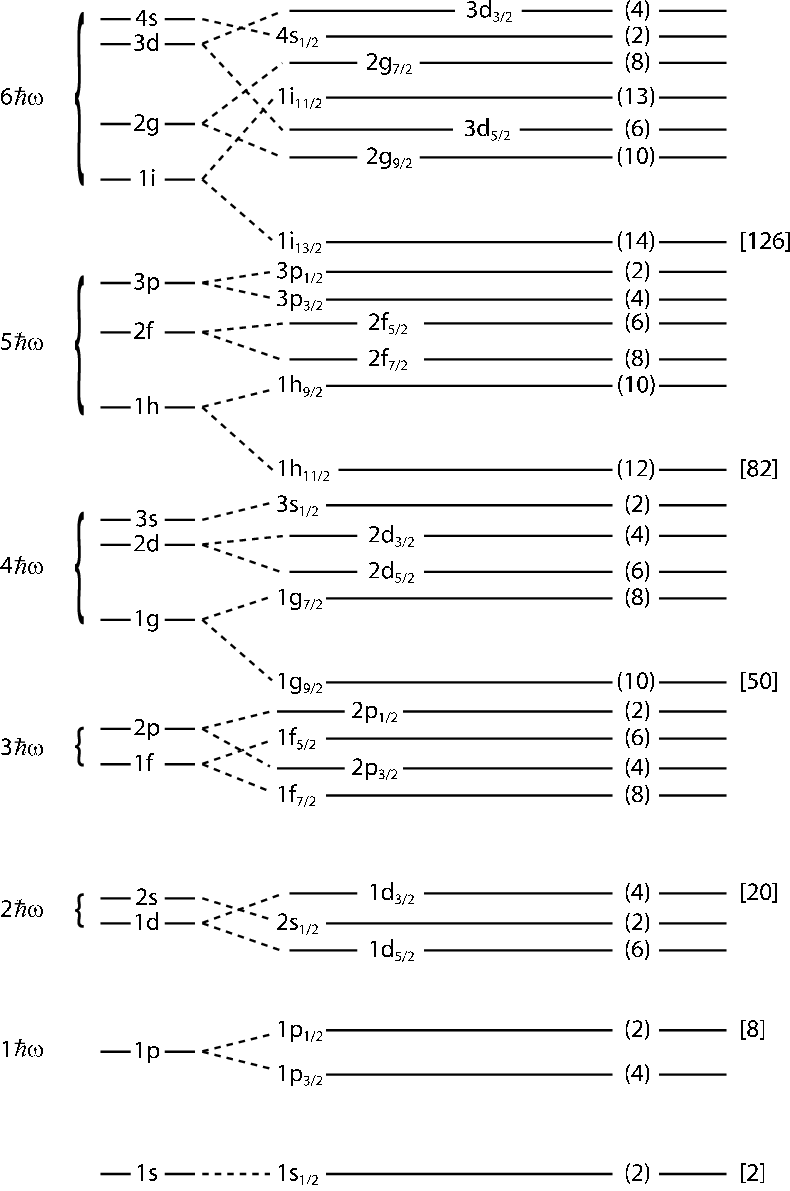}
\caption{Sequence of single particle states in the nuclear shell model.} 
\label{shellmod}
\end{center}
\end{figure}

The key to the Goeppert Mayer-Jensen success was the spin-orbit splitting, as
it turned out. There were the ``magic numbers,'' the large binding energies of
$^{16}$O, $^{40}$Ca, and $^{208}$Pb. Especially lead, with 82 protons and 126
neutrons, is very tightly bound.  Now it turns out (see Fig.\ \ref{shellmod})
that with the
strong spin-orbit splitting the $1h_{11/2}$ level for protons and the
$1i_{13/2}$ level for neutrons lie in both cases well below the next highest
levels. The notation here is different from that used in atomic physics. The
``1'' denotes that this is the first time that an $h$ ($l=5$) level would be
filled in adding protons to the nucleus and for the neutron levels similarly,
$i$ denoting $l=6$.

So how could a model in which nucleons move around in a common potential
without hitting each other be reconciled with the previous ``nuclear
porridge'' of Niels Bohr? Most of the answer came that the thorough mixture of
particles in the porridge arose because the neutron was dropped into the
nucleus with an energy of around 8 MeV above the ground state energy in the
cases of ``porridge.'' At least 
initially the shell model was classifying ground states of systems, where they
acquired no energy. The ground state was constructed by filling the lowest
energy single-particle states, one by one. Later, in 1958, Landau
formulated his theory of Fermi liquids, showing that as a particle (fermion)
was added with energy just at the top of the highest energy occupied state
(just at the top of the Fermi sea), the added particle would travel forever
without exciting the other particles. In more physical terms, the mean free
path (between collisions) of the particle is proportional to the inverse of
the square of the difference of its momentum from that of the Fermi surface;
i.e.,
\begin{equation}
\lambda \sim \frac{C}{(k-k_F)^2},
\label{mfp}
\end{equation}
where $C$ is a constant that depends on the interaction. Thus, as $k
\rightarrow k_F$, $\lambda \rightarrow \infty$ and the particle never 
scatters. Here $k_F$, the Fermi momentum, is that of the last filled orbit. 

Once the surprise had passed that one could assign a definite shell model
state to 
each nucleon and that these particles moved rather freely, colliding
relatively seldomly, the obvious question was how the self-consistent
potential the particles moved in could be constructed from the interactions
between the particles. The main technical problem was that these forces were
very 
strong. Indeed, Jastrow \cite{jast} characterized the short range force
between two nucleons as a vertical hard core of infinite height and radius of
$0.6\times 10^{-13}$ cm, about one-third the average distance between two
nucleons in nuclei. (Later, the theoretical radius of the core shrank to
$0.4\times 10^{-13}$ 
cm.) The core was much later found to be a rough characterization of the
short-range repulsion from vector meson exchange.

Now the whole postwar development of quantum electrodynamics by Feynman and
Schwinger was in perturbation theory, with expansion in the small parameter
$\alpha = e^2/ \hbar c = 1/137$.  Of course, infinities were encountered, but
since they shouldn't 
be there, they were set to zero. The concept of a hard core potential of
finite range, the radius a reasonable fraction of the average distance between
particles, was new. Lippmann and Schwinger \cite{lipp} had already
developed a formalism that could deal with such an interaction. In
perturbation theory; i.e., expansion of an interaction which is weak relative
to other quantities, the correction to the energy resulting from the
interaction is obtained by integrating the perturbing potential between the
wave functions; i.e.,
\begin{equation}
\Delta E \simeq \int \psi_0^\dag(x,y,z)\, V(x,y,z)\, \psi_0(x,y,z) \, dx\,
dy\, dz. 
\label{delE}
\end{equation}
Here $\psi_0(x,y,z)$ is the solution of the Schr\"{o}dinger wave equation for
the zero-order problem, that with $V(x,y,z) = 0$, and $\psi_0^\dag$ is the
complex conjugate of $\psi_0$. The wavefunction gives the probability of
finding a particle located in a small region $dx\, dy\, dz$ about the point
$(x,y,z)$:
\begin{equation}
P(x,y,z)=\psi_0^\dag (x,y,z)\, \psi_0(x,y,z)\, dx\, dy\, dz.
\end{equation}
The integral in eq.\
(\ref{delE}) is carried out over the entire region in which $V(x,y,z)$ is
nonzero.

Now we see the difficulty that arises if $V$ is infinite, as in the hard core
potential; namely, the product of an infinite $V$ and finite $\psi$ is
infinite. Eq.\ (\ref{delE}) is just the first-order correction to the
energy. More generally, perturbation theory yields a systematic expansion 
for the energy shift in terms of integrals involving higher powers of the
interaction ($V^2, V^3, \cdots$). As long as $\psi(x,y,z)$ is nonzero, all
of these terms are infinite.

Watson \cite{wat1,wat2} realized that a repulsion of any strength, even an
infinitely high hard core, could be handled by the formalism of Lippmann and
Schwinger, which was invented in order to handle two-body scattering. Quantum
mechanical scattering is described by the $T$-matrix:
\begin{equation}
T=V + V\frac{1}{E-H_0}V + V\frac{1}{E-H_0}V \frac{1}{E-H_0}V + \cdots, 
\label{Tmatser}
\end{equation}
where $V$ is the two-body potential, $E$ is the unperturbed energy, and $H_0$
is the unperturbed Hamiltonian, containing only the kinetic energy. This
infinite number of terms could be summed to give the result
\begin{equation}
T = V + V\frac{1}{E-H_0}T.
\label{Tmat}
\end{equation}
One can see that this is true by rewriting eq.\ (\ref{Tmatser}) as
\begin{equation}
T = V + V\frac{1}{E-H_0}\left (V + V\frac{1}{E-H_0}V +  V\frac{1}{E-H_0}V
  \frac{1}{E-H_0}V + \cdots \right),
\end{equation}
where the term in parentheses is clearly just $T$. Of course, Lippmann and
Schwinger did this with the proper mathematics, but the result eq.\
(\ref{Tmat}) came out of this. Watson realized
that the $T$-matrix made sense also for an extremely
strong repulsive interaction in a system of many nucleons. An incoming
particle could be scattered off each of the nucleons, one by one, and the
scattering amplitudes could be added up, the struck nucleon being left in the
same state as it was initially. The sum of the amplitudes could be squared to
give the total amplitude for the scattering off the nucleus.

Keith Brueckner, who was a colleague of Watson's at the University of Indiana
at the time, saw the usefulness of this technique for the nuclear many-body
problem. He was the first to recognize that the strong short-range
interactions, such as the infinite hard core, would scatter two nucleons to
momenta well above those filled in the Fermi sea. Thus, the exclusion
principle would have little effect and could be treated as a relatively small
correction.

Basically, for the many body problem,
the $T$-matrix is called the $G$-matrix, and the latter obeys the equation
\begin{equation}
G=V+V\frac{Q}{e} G.
\label{Gmat}
\end{equation}
Whereas in the Lippmann-Schwinger formula eq.\ (\ref{Tmat}) the energy
denominator $e$ was taken to be $E-H_0$, there was considerable debate about
what to put in for $e$ in eq.\ (\ref{Gmat}). We will see later that it is most
conveniently chosen to be $E-H_0$, as in eq.\ (\ref{Tmat}).\footnote{This
  conclusion will 
  be reached only after many developments, as we outline in Section 3.}
In eq.\ (\ref{Gmat}), the operator $Q$ excludes from the intermediate states
not only all of the 
occupied states below the Fermi momentum $k_F$, because of the Pauli
principle, but also all states beyond a 
maximum momentum $k_{max}$. The states below $k_{max}$ define what
is called the ``model space,'' which can generally be
chosen at the convenience of the investigator. The basic idea is that the
solution eq.\ (\ref{Gmat}) of $G$ is to be used as an effective interaction in
the space spanned by the Fermi momentum $k_F$ and $k_{max}$. This effective
interaction $G$ is then to be
diagonalized within the model space; i.e., the problem in that space with
the effective interaction $G$ is to be solved more or less exactly.

\begin{figure}
\begin{center}
\includegraphics[height=2in]{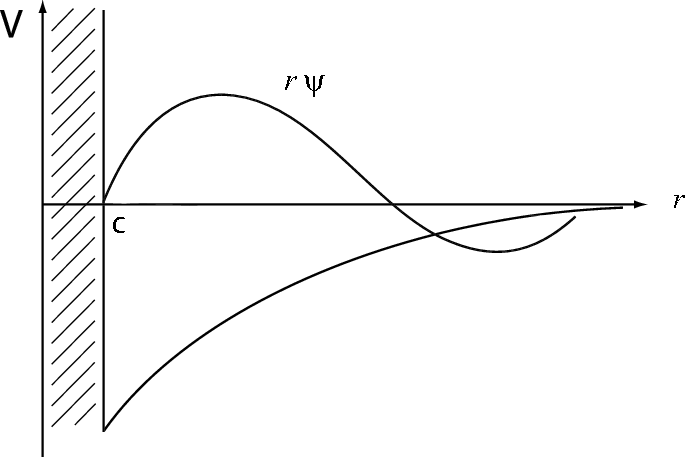}
\caption{The wave function $\psi$ for the relative motion of two particles
  interacting via a potential with an infinite hard core of radius $c$ and an
  attractive outside potential. It is convenient to deal with $r\psi$ rather
  than $\psi$.}
\label{wavefnt}
\end{center}
\end{figure}

In the above discussion leading to eq.\ (\ref{Gmat}) we have written the
nucleon-nucleon interaction as $V$. In fact, the interaction is complicated,
involving various combinations of spins and angular momenta of the two
interacting nucleons. In the time of Brueckner and the origin of his theory, it
took a lot of time and energy just to keep these combinations straight. This
large amount of bookkeeping is handled today fairly easily with electronic
computers. The major problem, however, was how to handle the strong
short-range repulsion, and this problem was discussed in terms of the
relatively simple, what we call ``central,'' interaction shown in Fig.\
\ref{wavefnt}. In fact, the $G$ of eq.\ (\ref{Gmat}) is still calculated from
that equation today in the most successful effective nuclear forces. The
$k_{max}$ is taken to be the maximum momentum at which experiments have been
analyzed, $k_{max} = 2.1 \mbox{ fm}^{-1} = \Lambda$, where $\Lambda$ is now
interpreted as a cutoff. Since experiments of momenta 
higher than $\Lambda$ have not been carried out and analyzed, at least not in
such a way as to bear directly on the determination of the potential, one
approach is to leave them out completely. The only
important change since the 1950's is that the $V(r)$, the first term on the
right hand side of eq.\ (\ref{Gmat}) is now rewritten in terms of a sum over
momenta, by what is called a Fourier transform, and this sum is truncated at
$\Lambda$, the higher momenta being discarded. The resulting effective
interaction, which replaces $G$, is now called $V_{\rm low-{\it k}}$.
\cite{bogner, meeft} We expand on this discussion at the end of this article.

The $G$-matrix of Brueckner was viewed by nuclear physicists as a complicated
object. However, it is clear what the effect of a repulsion at core radius $c$
that rises to infinite height will be--it will stop the two interacting
particles from going into the region of $r<c$. In non-relativistic quantum
mechanics, this means that their wave function of relative motion must be
zero. In other words, the wave function, whose square gives the probability of
finding the particle in a given region, must be zero inside the hard
core. Also, the wave function must be continuous outside, so that it must
start from zero at $r=c$. Therefore, we know that the wave function must look
something like that shown in Fig.\ \ref{wavefnt}. In any case, given the
boundary condition $\psi = 0$ at $r=c$, and the potential energy $V$, the wave
equation can be solved 
for $r>c$. It is not clear at this point how $V(r)$ is to be determined and
what the quantity $\psi$ is. We put off further discussion of this until
Section 3, in which we develop Hans Bethe's ``Reference Spectrum.''

\begin{figure}
\begin{center}
\includegraphics[height=3.5in]{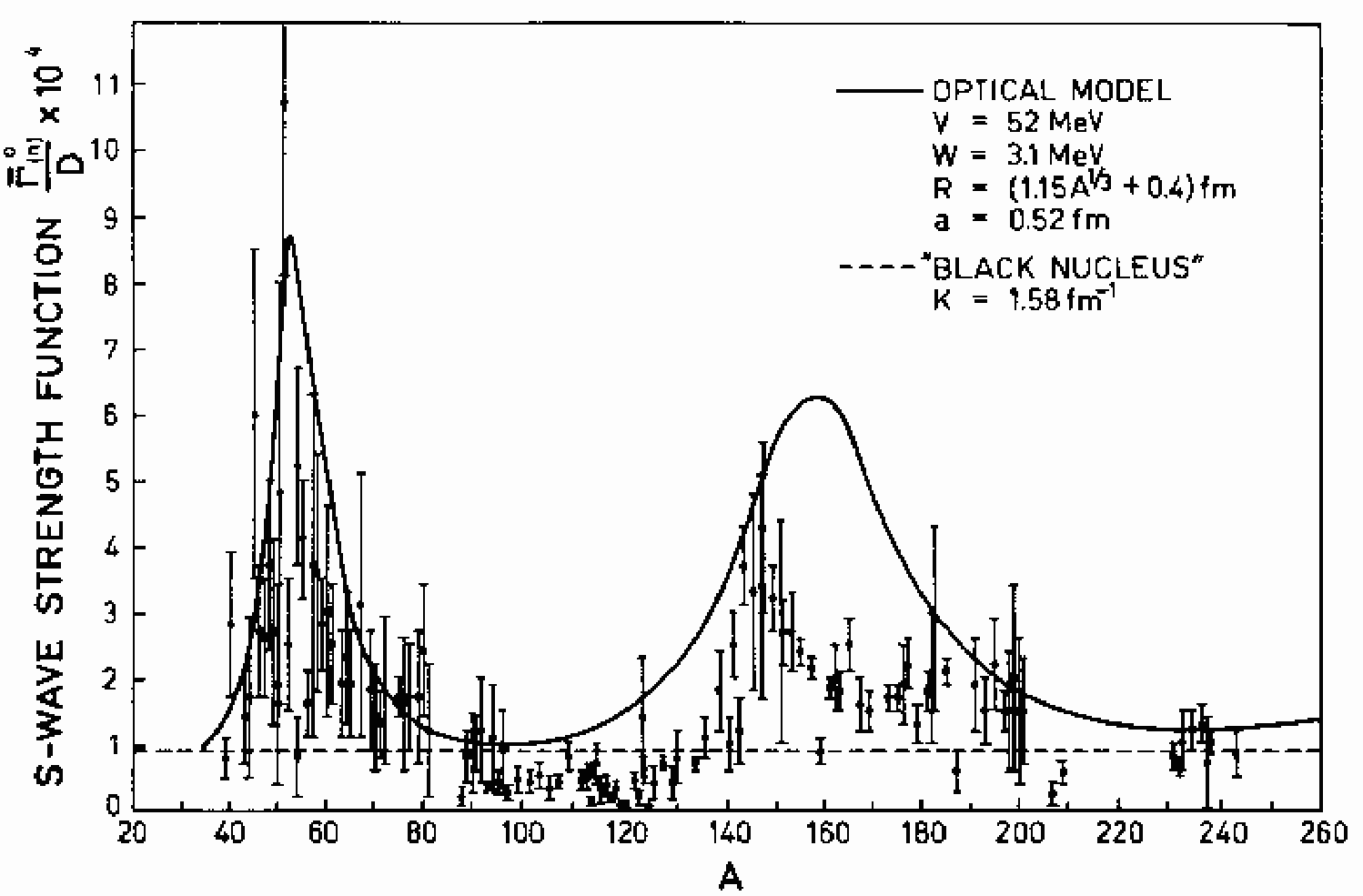}
\caption{The $S$-wave neutron strength function as a function of mass
  number. This figure is taken from A.\ Bohr and B.\ R.\ Mottelson, {\it
  Nuclear Structure}, Vol.\ 1, p.\ 230, W.\ A.\ Benjamin, New York, 1969.} 
\label{strfnt}
\end{center}
\end{figure}

One of the most important, if not the most important, influences on Hans Bethe
in his efforts, which we shall describe in the next section, to give a basis
for the nuclear shell model was the work of Feshbach, Porter, and Weisskopf
\cite{fpw}. These authors showed that although the resonances formed by
neutrons scattered by nuclei were indeed very narrow, their strength function
followed the envelope of a single-particle potential; i.e., of the strength
function for a single neutron in a potential $V+iW$. The strength function for
the compound nucleus resonances is defined as $\bar{\Gamma}_n/D$, were
$\Gamma_n$ is the width of the resonance for elastic neutron scattering
(scattering without energy loss) and $D$ is the average spacing between
resonances. This function gives the strength of 
absorption, averaged over many of the resonances. Parameters of the one-body
potential are given in the caption to Fig.\ \ref{strfnt}. The parameter $a$ is
the surface thickness of the one-body potential.

The peaks in the neutron strength function occur at those mass numbers where
the radius of the single-particle potential is just big enough to bind another
single-particle state with zero angular momentum. Although the single-particle
resonance is split up into the many narrow states discussed by Bohr, a vestige
of the single-particle shell model resonance still remains.

By contrast, an earlier literal interpretation of Bohr's model was worked out
by Feshbach, Peaslee, and Weisskopf in which the neutron is simply absorbed as
it enters the nucleus. This curve is the dashed line called ``Black Nucleus''
in Fig.\ \ref{strfnt}. It has no structure and clearly does not describe the
variations in the averaged neutron strength function. The Feshbach, Porter, 
and Weisskopf paper was extremely important in showing that there was an
underlying single-particle shell model structure in the individually extremely
complicated neutron resonances.

In fact, the weighting function $\rho(E-E^\prime)$ used by Feshbach, Porter,
and Weisskopf to calculate the average of the scattering amplitude $S(E)$
\begin{equation}
\langle S(E) \rangle _{Av} =
\int_{-\infty}^{\infty}\rho(E-E^\prime)F(E^\prime)dE^\prime,
\end{equation}
where $F(E)$ is an arbitrary (rapidly varying) function of energy $E$, was a
square one that had end effects which 
needed to be thrown away. A much more elegant procedure was suggested by Jim
Langer (see Brown \cite{brown}), which involved using the weighting function
\begin{equation}
\rho(E-E^\prime)=\frac{I}{\pi}\frac{1}{(E-E^\prime)^2+I^2}
\end{equation}
With this weighting function, the average scattering amplitude was
\begin{equation}
\left\{ \sum_n \frac{\Gamma(n)}{W_n-E} \right\}_{Av} =
\sum_n\frac{\Gamma(n)}{W_n-E-iI},
\end{equation}
where $W_n$ are the energies of the compound states with eV widths. Since the
$W_n$ are complex numbers all lying in the lower half of the complex plane, the
evaluation of the integral is carried out by contour integration, closing the
contour about the upper half plane.

Now if the $I$ is chosen to be about equal to the widths of the single-neutron
states in the optical model $V+iW$; i.e., of the order of $W=3.1$ MeV, then
the imaginary part of this average can be obtained as
\begin{equation}
Im\left(\left\{\sum_n \frac{\Gamma(n)}{W_n-E} \right\}_{Av}\right) =
\pi\frac{\bar{\Gamma}(n)}{D} = Im\sum_m
\left(\frac{\Gamma_m}{\hat{E}_m-E}\right), 
\end{equation}
where $\Gamma_m$ and $\hat{E}_m$ are the widths and (complex) energies of the
single neutron states in the complex potential $V+iW$. This shows how the
averaged strength function is reproduced by the single-particle levels.

In the summer of 1958 Hans Bethe invited me (G.E.B.) to Cornell, giving me an
honorarium from a fund that the AVCO Company, for which he consulted on the
physics of the nose cones of rockets upon reentry into the atmosphere, had
given him for that purpose. I was unsure of the convergence of the procedure
by which I had obtained the above results. Hans pointed out that the width of
the single-neutron state $Im(\Gamma_m)$ would be substantially larger than the
widths of the two-particle, one-hole states that the single-particle state
would decay into, because in the latter the energy would be divided into the
three excitations, two particles and one hole, and the widths went
quadratically with available energy as can be obtained from eq.\ (\ref{mfp});
ergo the two-particle, one-hole widths would be down by a factor of $\sim 3/9
= 1/3$ from the single-particle width, but nonetheless would acquire the same
imaginary part $I$, which is the order of the single particle width in the
averaging. Thus, my procedure would be convergent. So I wrote my {\it Reviews
  of Modern Physics} article \cite{brown}, which I believe was quite
elegant, beginning with Jim Langer's idea about averaging and ending with Hans
Bethe's argument about convergence. I saw then clearly the advantage of having
good collaborators, but had to wait two more decades until I could lure Hans
back into astrophysics where we would really collaborate tightly.

\section{Hans Bethe in Cambridge, England}

We pick up Hans Bethe at the time of his sabbatical in Cambridge, England. The
family, wife Rose, and children, Monica and Henry, were with him.

There was little doubt that Keith Brueckner had a promising approach to attack
the nuclear many body problem; i.e., to describe the interactions between
nucleons in such a way that they could be collected into a general
self-consistent potential. That potential would have its conceptual basis in
the 
Hartree-Fock potential and would turn out to be the shell model potential,
Hans realized.

Douglas Hartree was professor at the University of Manchester when he invented
the self-consistent Hartree fields for atoms in 1928. He put $Z$ electrons
into wave functions around a nucleus of $Z$ protons, $A-Z$ neutrons, the
latter having no effect because they had no charge. The heavy compact nucleus
was taken to be a point charge at the origin of the coordinate system, because
the nucleon mass is nearly 2000 times greater than the electron and the size
of the nucleus is about 1000 times smaller in radius than that of the
atom. The nuclear 
charge $Ze$ is, of course, screened by the electron charge as electrons
gather about it. The two innermost electrons are very accurately in $1s$
orbits (called $K$-electrons in the historical nomenclature). Thus, the other
$Z-2$ electrons see a screened charge of ($Z-2$)$e$, and so it could go, but
Hartree used instead the so-called Thomas-Fermi method to get a beginning
approximation to the screened electric field.

Given this screened field as a function of distance $r$, measured from the
nucleus located at $r=0$, Hartree then sat down with his mechanical computer
punching buttons as his Monromatic or similar machine rolled back and forth,
the latter as he hit a return key. (\mbox{G.E.B.--These} machines tore at my
eyes, giving me 
headaches, so I returned to analytical work in my thesis in 1950. Hans used a
slide rule, whipping back and forth faster than the Monromatic could travel,
achieving three-figure accuracy.) When Hartree had completed the solution of
the 
Schr\"{o}dinger equation for each of the original $Z$ electrons, he took its
wave function and squared it. This gave him the probability,
$\rho_i(x_i,y_i,z_i)$, of finding electron $i$ at the position given by the
coordinates $x_i,y_i,z_i$. Then, summing over $i$, with $x_i=x$, $y_i=y$, and
$z_i=z$
\begin{equation}
\sum_i\rho_i(x,y,z)=\rho_e^{(1)}(x,y,z)
\end{equation}
gave him the total electron density at position ($x$, $y$, $z$). Then he began
over again with a new potential
\begin{equation}
V^{(1)}=\frac{Ze}{r}-e\rho_e^{(1)}(x,y,z),
\end{equation}
where superscript (1) denotes that this is a first approximation to the
self-consistent potential. To reach approximation (2) he repeated the process,
calculating the $Z$ electronic wave functions by solving the Schr\"{o}dinger
equation with the potential $V^{(1)}$. This gave him the next Hartree
potential $V^{(2)}$. He kept going until the potential no longer changed upon
iteration; i.e., until  $V^{(n+1)} \simeq V^{(n)}$, the $\simeq$ meaning that
they were approximately equal, to the accuracy Hartree desired. Such a
potential is called ``self-consistent'' because it yields an electron density
that reproduces the same potential.

Of course, this was a tedious job, taking months for each atom (now only
seconds with electronic computers). Some of the papers are coauthored, D.\ R.\
Hartree and W.\ Hartree. The latter Hartree was his father, who wanted to
continue working after he retired from employment in a bank. In one case,
Hartree made a mistake in transforming his units $Z$ and $e$ to more
convenient dimensionless units, and he performed calculations with these
slightly incorrect
units for some months. Nowadays, young investigators might nonetheless try to
publish the results as referring to a fractional $Z$, hoping that fractionally
charged particles would attach themselves to nuclei, but Hartree threw the
papers in the wastebasket and started over.

Douglas Hartree was professor in Cambridge in 1955 when Hans Bethe went there
to spend his sabbatical. The Hartree method was improved upon by the Russian
professor Fock, who added the so-called exchange interaction which enforced
the Pauli exclusion principle, guaranteeing that two electrons could not occupy
the same state. We shall simply enforce this principle by hand in the
following discussion, our purpose here being to explain what
``self-consistent'' means in the many-body context. Physicists generally
believe self-consistency to be a good attribute of a theory, but Hartree did
not have to 
base his work only on beliefs. Given his wave functions, a myriad of
transitions between atomic levels could be calculated, and their energies and
probabilities could be compared with experiment. Douglas Hartree became a
professor at Cambridge. This indicates the regard in which his work was held.

The shell model for electrons reached success in the Hartree-Fock
self-consistent field approach. This was very much in Hans Bethe's mind, when
he set out to formulate Brueckner theory so that he could obtain a
self-consistent potential for nuclear physics. He begins his 1956 paper 
\cite{bethe1} on the ``Nuclear Many-Body Problem'' with ``Nearly everybody in
nuclear physics has marveled at the success of the shell model. We shall use
the expression `shell model' in its most general sense, namely as a scheme in
which each nucleon is given its individual quantum state, and the nucleus as a
whole is described by a `configuration,' i.e., by a set of quantum numbers for
the individual nucleons.''

He goes on to note that even though Niels Bohr had shown that low-energy
neutrons disappear into a ``porridge'' for a very long time before re-emerging
(although Hans was not influenced by Bohr's paper, because he hadn't read
it; this may have given him an advantage as we shall see), Feshbach, Porter,
and Weisskopf had shown that the envelope of these states followed that of the
single particle state calculated in the nuclear shell model, as we noted in
the last section.

Bethe confirms ``while the success of the model [in nuclear physics]
has thus been beyond question for many years, a theoretical basis for it has
been lacking. Indeed, it is well established that the forces between two
nucleons are of short range, and of very great strength, and possess exchange
character and probably repulsive cores. It has been very difficult to see how
such forces could lead to any over-all potential and thus to well-defined
states for the individual nucleons.''

He goes on to say that Brueckner has developed a powerful mathematical method
for calculating the nuclear energy levels using a self-consistent field
method, even though the forces are of short range. 

``In spite of its apparent great accomplishments, the theory of Brueckner {\it 
  et al.} has not been readily accepted by nuclear physicists. This is in large
measure the result of the very formal nature of the central proof of the
  theory. In addition, the 
definitions of the various concepts used in the theory are not always
clear. Two important concepts in the theory are the wave functions of the
individual particles, and the potential $V_c$ `diagonal' in these states. The
paper by Brueckner and Levinson \cite{brulev} defines rather clearly how the
potential is to be obtained from the wave functions, but not how the wave
functions can be constructed from the potential $V_c$. Apparently, BL assume
tacitly that the nucleon wave functions are plane waves, but in this case, the
method is only applicable to an infinite nucleus. For a finite nucleus, no
prescription is given for obtaining the wave functions.''

Hans then goes on to define his objective, which will turn out to develop into
his main activity for the next decade or more: ``It is the purpose of the
present paper to show that the theory of Brueckner gives indeed the foundation
of the shell model.''

Hans was rightly very complimentary to Brueckner, who had ``tamed'' the
extremely strong short-ranged interactions between two nucleons, often taken
to be infinite in repulsion at short distances at the time. On the other hand,
Brueckner's immense flurry of activity, changing and improving on previous
papers, made it difficult to follow his work. Also, the Watson input
scattering theory seemed to give endless products of scattering operators,
each appearing to be ugly mathematically (``Taming'' thus was the great
accomplishment of Watson and Brueckner). And in the end, the real goal was to
provide a 
quantitative basis for the nuclear shell model, based on the nucleon-nucleon
interaction, which was being reconstructed from nucleon-nucleon scattering
experiments at the time.

The master at organization and communication took over, as he had done in
formulating the ``Bethe Bible'' during the 1930's.

One can read from the Hans Bethe archives at Cornell that Hans first made 100
pages of calculations reproducing the many results of Brueckner and
collaborators, before he began numbering his own pages as he worked out the
nuclear many body problem. He carried out the 
calculations chiefly analytically, often using mathematical functions,
especially spherical Bessel functions, which he had learned to use while with
Sommerfeld. When necessary, he got out his slide rule to make numerical
calculations.

During his sabbatical year, Hans gave lectures in Cambridge on the nuclear
many body problem. Two visiting American graduate students asked most of the
questions, the British students being reticent and rather shy. But Professor
Neville Mott asked Hans to take over the direction of two graduate
students, Jeffrey Goldstone and David Thouless, the former now professor at
M.I.T. and the latter professor at the University of Washington in Seattle. We
shall return to them later.

As noted earlier, the nuclear shell model has to find its own center
``self-consistently.'' Since it is spherically symmetrical, this normally
causes no 
problem. Simplest is to assume the answer: begin with a deep square well or
harmonic oscillator potential and fill it with single particle eigenstates as
a zero-order approximation as Bethe and Bacher \cite{bethbach} did in their
1936 paper of the Bethe Bible. Indeed, now that the sizes and shapes of nuclei
have been measured by high resolution electron microscopes (high energy
electron scattering), one can reconstruct these one-particle potential wells
so that filling them with particles reproduces these sizes and shapes. The wave
functions in such wells are often used as assumed solutions to the
self-consistent potential that would be obtained by solving the nuclear
many-body problem. 

We'd like to give the flavor of the work at the stage of the Brueckner theory,
although the work using the ``Reference-Spectrum'' by Bethe, Brandow,
and Petschek \cite{bbp}, which we will discuss later, will be much more
convenient for understanding the nuclear shell model. The question we consider
is the magnitude of the three-body cluster terms; i.e., the contribution to the
energy from the interaction of three particles. (Even though the elementary
interaction may be only a two-particle one, an effective three-body
interaction arises inside the nucleus, as we shall discuss.) One particular
three-body cluster will be shown in Fig.\ \ref{threebod}. The three-body
term, called a three-body cluster in the Brueckner expansion, is
\begin{equation}
\Delta E_3 = \sum_{ijk}\left\langle \Phi_0 \left|
  I_{ij}\frac{Q}{e_{ji}}I_{jk}\frac{Q}{e_{ki}}I_{ki} \right| \Phi
  \right\rangle. 
\label{tbc}
\end{equation}
Here the three nucleons $i$, $j$, and $k$ are successively excited. There are,
of course, higher order terms in which more nucleons are successively excited
and de-excited. In this cluster term, working from right to left, first there
is an interaction $I_{ki}$ between particles $i$ and $k$. In fact, this pair
of particles is allowed to interact any number of times, the number being
summed into the $G$-matrix $G_{ik}$. We shall discuss the $G$-matrix in great
detail later. Then the operator $Q$ excludes all intermediate states that are
occupied by other particles--the particle $k$ can only go into an unoccupied
state before interacting with particle $j$. The denominator $e_{ki}$ is the
difference in energy between the specific state $k$ that particle $i$ goes to
and the state that it came from. The particle $k$ can make virtual
transitions, transitions that don't conserve energy, because of the Heisenberg
uncertainty principle
\begin{equation}
\Delta E \Delta t \gtrsim \hbar.
\end{equation}
Thus, if $\Delta E \Delta t > \hbar$, then the particle can stay in a given
state only a time
\begin{equation}
t=\hbar/e_{ki}
\end{equation}
so the larger the $e_{ki}$ the shorter the time that the particle in a given
state can contribute to the energy $\Delta E_3$. (Of course, the derivation of
$\Delta E_3$ is carried out in the standard operations of quantum
mechanics. We bring in the uncertainty principle only to give some
qualitative understanding of the result.) Once particle $k$ has interacted
with particle $j$, particle $j$ goes on to interact with the original particle
$i$, since the nucleus must be left in the same ground state $\Phi_0$ that it
began in, if the three-body cluster is to contribute to its energy.

Bethe's calculation gave
\begin{equation}
\Delta E_3/A = -0.66 \mbox{ MeV}
\end{equation}
corrected a bit later in the paper to $-0.12$ MeV once only the fraction of
spin-charge states allowed by selection rules are included. This is to be
compared with Brueckner's $-0.007$ MeV. Of course, these are considerably
different, but this is not the main point. The main point is that both
estimates are small compared with the empirical nuclear binding
energy.\footnote{At least at that time there appeared to be a good convergence 
  in the so-called cluster expansion. But see Section 3!}

Thus, it was clear that the binding energy came almost completely from the
two-body term $G_{ik}$, and that the future effort should go into evaluating
this quantity, which satisfies the equation
\begin{equation}
G=V+V\frac{Q}{e}G.
\end{equation}
(We will have different $G$'s for the different charge-spin states.)

Although written in a deceptively simple way, this equation is ugly, involving
operators in both the coordinates $x$, $y$, $z$, and their
derivatives. However, $G$ can be expressed as a two-body operator (see Section
3). It does not involve a sum over the other particles in the nucleus, so the
interactions can be evaluated one pair at a time.

Thus, the first paper of Bethe on the nuclear many body problem collected the
work of Brueckner and collaborators into an orderly formalism in which the
evaluation of the two-body operators $G$ would form the basis for calculating
the shell model potential $V(r)$. Bethe went on with Jeffrey Goldstone to
investigate the evaluation of $G$ for the extreme infinite-height hard core
potential, and he gave David Thouless the problem that, given the empirically
known shell model potential $V_{SM}(r)$, what properties of $G$ would
reproduce it.

As we noted in the last section, an infinite sum over the two-body interaction
is needed in order to completely exclude the two nucleon wave functions from
the region inside the (vertical) hard core. As noted earlier, Jastrow
\cite{jast} had proposed a vertical hard core, rising 
to $\infty$, initially of radius $R_c = 0.6 \times 10^{-13}$ cm. This is about
1/3 of the distance between nucleons, so that removing this amount of space
from their possible occupancy obviously increases their energy. (From the
Heisenberg Principle $\Delta p_x \Delta x \gtrsim \hbar$, which is commonly
used to 
show that if the particles are confined to a smaller amount of volume, the
$\Delta x$ is decreased, so the $\Delta p_x$, which is of the same general
size as $p_x$, is increased.) Thus, as particles were pushed closer together
with increasing density, their energy would be greater. Therefore, the
repulsive core was thought to be a great help in saturating matter made up of
nucleons. One of the authors (G.E.B.) heard a seminar by
Jastrow at Yale in 1949, and Gregory Breit, who was the leading theorist
there, thought well of the idea. Indeed, we shall see later that Breit played
an important role in providing a physical mechanism for the hard core as we
shall discuss in the next section. This mechanism actually removed the ``sharp
edges'' which gave the vertical hard core relatively extreme properties. So we
shall see later that the central hard core in the interaction between two
nucleons isn't vertical, rather it's of the form
\begin{equation}
\mbox{Hard Core } =\frac{C}{r}\mbox{ exp}\left( -\frac{m_\omega r}{\hbar
    c}\right), 
\end{equation}
where $\hbar/m_\omega c$ is $0.25 \times 10^{-13}$ cm, and $m_\omega$ is the
mass of the $\omega$-meson. The constant $C$ is large compared with unity
\begin{equation}
C \gg 1
\end{equation}
so that the height of the core is many times the Fermi energy; i.e., the
energy measured from the bottom of the shell model potential well up to the
last filled level. Thus, although extreme, treating the hard core potential
gives a caricature problem. Furthermore, solving this problem in ``Effect of
a repulsive core in the theory of complex nuclei,'' by H.\ A.\ Bethe and J.\
Goldstone \cite{bg} gave an excellent training to Jeffrey Goldstone, who went
on to even greater accomplishments. (The Goldstone boson is named after him;
it is perhaps the most essential particle in QCD.)

We will find in the next section that the Moszkowski-Scott separation method
is a more convenient way to treat the hard core. And this method is more
practical because it includes the external attractive potential at the same
time. However, even though the wave function of relative motion of the two
interacting particles cannot penetrate the hard core, it nonetheless has an
effect on their wave function in the external region. This effect is
conveniently included by changing the reduced mass of the interacting nucleons
from $M^\star$ to $0.85M^\star$ for a hard core radius $r_c = 0.5 \times
10^{-13}$ cm, and to $\sim 0.9M^\star$ for the presently accepted $r_c = 0.4
\times 10^{-13}$ cm.

\section{The Reference-Spectrum Method for Nuclear Matter}

We take the authors' prerogative to jump in history to 1963, to the paper of
H.\ A.\ Bethe, B.\ H.\ Brandow, and A.\ G.\ Petschek \cite{bbp} with the title
that of this section, because this work gives a convenient way of discussing
essentially all of the physical effects found by Bethe and collaborators, and
also the paper by S.\ A.\ Moszkowski and B.\ L.\ Scott \cite{moscot} the latter
paper being very important for a simple understanding of the $G$-matrix. In
any case, the reference-spectrum included in a straightforward way all of the
many-body effects experienced by the two interacting particles in Brueckner
theory, and enabled the resulting $G$-matrix to be written as a function of
$r$ alone, although separate functions $G_l(r)$ had to be obtained for each
angular momentum $l$.

Let us prepare the ground for the reference-spectrum, including also the
Moszkowski-Scott method, by a qualitative discussion of the physics involved in
the nucleon-nucleon interaction inside the nucleus, say at some reasonable
fraction of nuclear matter density. We introduce the latter so we can talk
about plane waves locally, as an approximation. For simplicity we take the
short-range repulsion to come from a hard core, of radius $0.4 \times
10^{-13}$ cm.

\begin{figure}
\begin{center}
\includegraphics[height=2in]{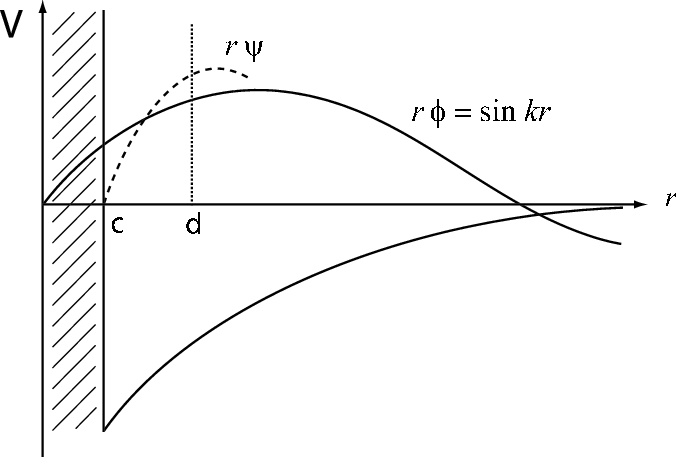}
\caption{The unperturbed $S$-wave function $r\phi = \mbox{sin } kr$ and the
  wave function $r\psi$ obtained by integrating the Schr\"{o}dinger equation
  in the presence of the potential out from $r=c$.} 
\label{sepdist}
\end{center}
\end{figure}

(i) In the region just outside the hard core, the influence of neighboring
nucleons on the two interacting ones is negligible, because the latter have
been kicked up to high momentum states by the strong hard-core interaction and
that of the attractive potential, which is substantially stronger than the
local Fermi energy of the nucleons around the two interacting ones. Thus, one
can begin integrating out the Schr\"{o}dinger equation for $r\psi (r)$,
starting from $r\psi(r) =0$ at $r=r_c$. The particles cannot penetrate the
infinitely repulsive hard core (see Fig.\ \ref{sepdist}), so their wave
function 
begins from zero there. In fact, the Schr\"{o}dinger equation is one of
relative motion; i.e., a one-body equation in which the mass is the reduced
mass of the two particles, $M_N/2$ for equal mass nucleons. As found by Bethe
and Goldstone, the $M_N$ will be changed to $M_N^\star$ in the many-body
medium, but the mass is modified not only by the hard core but also by the
attractive part of the two-body potential, which we have not yet discussed.

(ii) We consider the spin-zero $S=0$ and spin-one $S=1$ states; i.e., for
angular momentum $l=0$. These are the most important states. We compare the
spin-one state in the presence of the potential with the unperturbed one;
i.e., the one in the absence of a potential.

Before proceeding further with Hans Bethe's work, let us characterize the nice
idea of Moszkowski and Scott in the most simple possible way.

Choose the separation point $d$ such that
\begin{equation}
\left. \frac{d\psi(r)/dr}{\psi(r)}\right|_{r=d} =
\left. \frac{d\phi(r)/dr}{\phi(r)}\right|_{r=d} 
\label{sepdis}
\end{equation}
Technically, this is called the equality of logarithmic derivatives. As shown
in Fig.\ \ref{sepdist}, such a point will be there, because $r\psi$, although
it 
starts from zero at $r=c$, has a greater curvature than $r\phi$. Now if only
the potential inside of $d$ were present; i.e., if $V(r)=0$ for $r>d$, then
eq.\ (\ref{sepdis}) in quantum mechanics is just the condition that the inner
potential would produce zero scattering, the inner attractive potential for
$r<d$ just canceling the repulsion from the hard core. Of course, if eq.\
(\ref{sepdis}) is true for a particular $k$, say $k\lesssim k_F$, then it will
not be exactly true for other $k$'s. On the other hand, since the momenta in
the short-distance wave function are high compared with $k_F$, due to the
infinite hard core, and the very deep interior part of the attractive
potential that is needed to compensate for it, eq.\ (\ref{sepdis}) is
nearly satisfied for all momenta up to $k_F$ if the equality is true for one
of the momenta.

The philosophy here is very much as in Bethe's work ``Theory of the Effective
Range in Nuclear Scattering'' \cite{bethe2}. This work is based on the fact
that the inner potential (we could define it as $V(r)$ for $r<d$) is deep in
comparison with the energy that the nucleon comes in with, so that the
scattering depends only weakly on this (asymptotic) energy. In fact, for a
potential which is Yukawa in nature
\begin{equation}
V=\infty, \mbox{ for } r<c=0.4 \times 10^{-13} \mbox{ cm}
\end{equation}
\begin{equation}
V=-V_0e^{-(r-c)/R} \mbox{ for } r>c,
\end{equation}
then $V_0 = 380$ MeV and $R=0.45\times 10^{-13}$ cm in order to fit the low
energy neutron-proton scattering \cite{brownjack}. The value of $380$ MeV is
large compared with the Fermi energy of $\sim 40$ MeV. In the collision of two
nucleons, the equality of logarithmic derivatives, eq.\ (\ref{sepdis}), would
mean that the inner part of the potential interaction up to $d$, which we call
$V_s$, would give zero scattering. All the scattering would be given by the
long-range part which we call $V_l$.

(iii) Now we know that the wave function $\psi(r)$ must ``heal'' to $\phi(r)$
as $r\rightarrow \infty$, because of the Pauli principle. There is no other
place for the particle to go, because for $k$ below $k_F$ all other states are
occupied. Delightfully simple is to approximate the healing by taking $\psi$
equal to $\phi$ for $r>d$.

The conclusion of the above is that
\begin{equation}
G(r) \simeq G_s + v_l(r) +v_l\frac{Q}{e}v_l + \cdots,
\label{msg}
\end{equation}
where 
\begin{equation}
v_l=V(r) \mbox{ for } r\gtrsim d.
\end{equation}
We shall discuss $G_s$, the $G$-matrix which would come from the short-range
part of the potential for $r<d$ later. It will turn out to be small.

Now we have swept a large number of problems under the rug, and we don't
apologize for it because eq.\ (\ref{msg}) gives a remarkably accurate
answer. However, first we note that there must be substantial attraction in the
channel considered in order for the short-range repulsion to be canceled. So
the separation method won't work for cases where there is little
attraction. Secondly, a number of many-body effects have been discarded, and
these had to be considered by Bethe, Brandow, and Petschek, one by one.

\begin{figure}
\begin{center}
\includegraphics[height=2in]{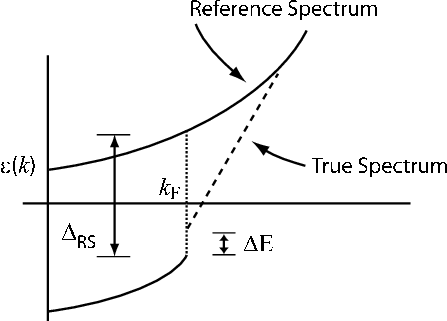}
\caption{Energy spectrum in the Brueckner theory (inclusive of self-energy
  insertion). The true spectrum has a small gap $\Delta E$ at $k_F$ and is
  known, from the arguments of Bethe {\it et al.}, to join the
  reference-spectrum at $k\approx 3$ fm$^{-1}$. The dotted line is an
  interpolation between these two points.} 
\label{refspec}
\end{center}
\end{figure}

We do not want to fight old battles over again, but simply note that Bethe,
Brandow, and Petschek found that they could take into account the many-body
effects including the Pauli principle by choosing a single-particle spectrum,
which is approximated by
\begin{equation}
\epsilon_m=\frac{k_m^2}{2m^\star_{RS}} + A.
\label{rssps}
\end{equation}
The parameters $m^\star_{RS}$ and $A$ were arrived at by a self-consistency
process. The hole energies; i.e., the energies of the initially bound
particles, can be treated fairly roughly since they are much smaller than the
energies of the particles which have quite high momenta (see Fig.\
\ref{refspec}). Given an input $m^\star_{RS}$ and $A$, the same $m^\star_{RS}$
and 
$A$ must emerge from the calculation. Thus, the many-body effects can be
included by changing the single-particle energy through the coefficient of the
kinetic energy and by adding the constant $A$. In this way the many-body
problem is reduced to the Lippmann-Schwinger two-body problem with changed
parameters, $m\rightarrow m^\star_{RS}$ and the addition of $A$. As shown in
Fig.\ \ref{refspec}, the effect is to introduce a gap $\Delta_{RS}$ between 
particle
and hole states.\footnote{We call the states initially in the Fermi sea ``hole 
  states'' because holes are formed when the two-body interaction transfers
  them to the particle states which lie above the Fermi energy, leaving holes
  behind in intermediate states.}

A small gap, shown as $\Delta E$, occurs naturally between particle and hole
states, resulting from the unsymmetrical way they are treated in Brueckner
theory, the particle-particle scattering being summed to all orders. The main
part of the reference-spectrum gap $\Delta_{RS}$ is introduced so as to
numerically reproduce the effects of the Pauli Principle. That this can be
done is not surprising, since the effect of either is to make the wave
function heal more rapidly to the noninteracting one.

Before we move onwards from the reference-spectrum, we want to show the main
origin of the reduction of $m$ to $m^\star_{RS} \sim 0.8m$. Such a reduction
obviously increases the energies of the particles in intermediate states
$\epsilon(k)$ which depends inversely on $m^\star_{RS}$. A lowering of
$m^\star$ from $m$ was already found in the Bethe-Goldstone solution of the
scattering from a hard core alone with $m^\star=0.85m$ for a hard core radius
of $r_c=0.5\times 10^{-13}$ cm and somewhat less for $r_c=0.4\times 10^{-13}$
cm, a more reasonable value. The $m^\star_{RS}$ are not much smaller than
this.

\begin{figure}
\begin{center}
\includegraphics[height=1.5in]{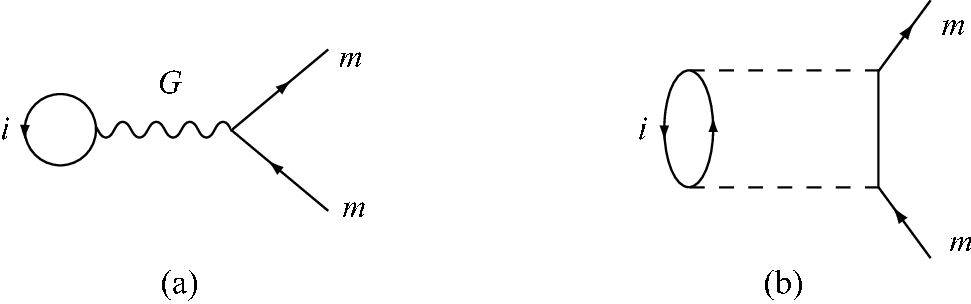}
\caption{(a) The on-shell self energy involves the interaction of a particle in
  state $m$ with the filled states in the Fermi sea $i$. The wiggly line $G$
  implies a sum over all orders in a ladder of $V$, the second-order term in
  dashed lines being shown in (b). Together with the kinetic energy, this
  gives the self energy.} 
\label{onshell}
\end{center}
\end{figure}

Now we have to introduce the concept of off-energy-shell self energies. Let's
begin by defining the on-shell self-energy which is just the energy of a
particle in the shell model or optical model potential. It would be given by
the process shown in Fig.\ \ref{onshell}.

\begin{figure}
\begin{center}
\includegraphics[height=1.5in]{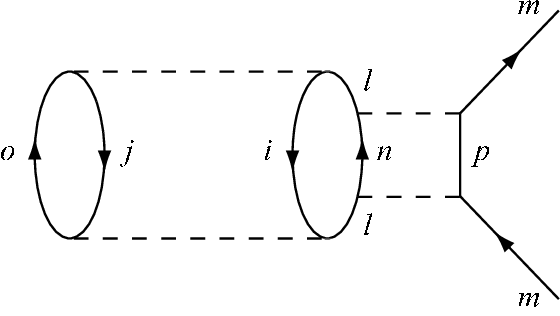}
\caption{Second-order contribution to the self energy of a particle in state
  $m$ when present in a process contributing to the ground-state energy. The
  states $l,m,n,o,p$ refer to particles and $i,j$ to holes.} 
\label{offshell}
\end{center}
\end{figure}

The on-shell $U(k_m)$ should just give the potential energy of a nucleon of
momentum $k$ in the optical model potential $U+iW$ used by Feshbach, Porter,
and Weisskopf, so $U(k_m) + k_m^2/2m$ gives the single-particle energy.

However, the energies $U(k_m)$ to be used in Brueckner theory are not on-shell
energies. This is because another particle must be excited when the one being
considered is excited, as shown in Fig.\ \ref{offshell}. Thus, considering the
second order in $V$ interaction between the particle being considered and
those in the nucleus one has
\begin{equation}
\tilde{\epsilon}^{(2)}_p = -\sum_{p,n>k_F;i<k_F}\frac{\left| \langle np \left|
V \right| lm\rangle \right|^2}{\epsilon_n+\epsilon_p+\epsilon_o-\epsilon_j-
\epsilon_i -\epsilon_m}
\label{so}
\end{equation}
i.e., although the particle in state $o$ and the hole in state $j$ do not
enter actively into the self energy of $p$, one must include their energies in
the energy denominator. Defining
\begin{equation}
\tilde{\epsilon}= \epsilon_n+\epsilon_p+\epsilon_o-\epsilon_j-
\epsilon_i -\epsilon_m,
\end{equation}
\begin{equation}
e=\epsilon_n+\epsilon_p-\epsilon_i -\epsilon_m
\end{equation}
one sees that the additional energy in the denominator in eq.\ (\ref{so}) is
\begin{equation}
\Delta e=\tilde{\epsilon}-\epsilon=\epsilon_o-\epsilon_j.
\end{equation}
This $\Delta e$ represents the energy that must be ``borrowed'' in order to
excite the particle from state $j$ to $o$, even though this particle does not
directly participate in the interaction on state $p$ which gives its self
energy. The point is that because $j$ must also be excited, as well as the
particles on the right in the diagram of Fig.\ \ref{offshell}, the total
$\tilde{e}$ 
is that much greater than the $e$ necessary to give the on-shell energy. Thus,
the time that this interaction can go on for is decreased, since, from the
uncertainty principle the time that energy can be borrowed for goes as $\Delta
t = \hbar/\Delta E$. As noted earlier, the wave function $\psi(r)$ lies outside
the hard core, so that integrals over the $G$-matrix are only over the
attractive interaction. These are cut down by the additional energy
denominator that comes in by the interaction being off shell. In the higher
momentum region this can cut the self energy down by 15--20 MeV. We shall see
later that when this is handled properly, only $\sim 1/3$ of this survives.

The strategy that T.\ T.\ S.\ Kuo and one of the authors (G.E.B.) have employed
over many years, beginning with Kuo and Brown \cite{kuobrown} to calculate the
$G$-matrix in nuclei has been to use the separation method for the $G$ in the
$l=0$ states ($S$-states), which have a large enough attraction so that a
separation distance $d$ can be defined, but to use the Bethe, Brandow, and
Petschek reference-spectrum in angular momentum states which do not have this
strong attraction.

The reference-spectrum was as far as one could get using only two-body
clusters; i.e., summing up the two-body interaction. In order to make further
progress, three and four-body clusters had to be considered; i.e., processes
in which one pair of the initially interacting particles was left excited
while another pair interacted, and only then returned to its initial state as
in the calculations of the lowest order three-body cluster $\Delta E_3$, eq.\
(\ref{tbc}).

Hans Bethe investigated the four-body cluster in Ref.\ \cite{bethe3}, following
work on three-nucleon clusters in nuclear matter by his post-doc R.\ (Dougy)
Rajaraman in Ref.\ \cite{raj}, a paper just following the Bethe, Brandow, and
Petschek reference-spectrum paper. Rajaraman suggested on the basis of
including the other three-body clusters than that shown in Fig.\
\ref{onshell}, that 
the off-shell effect should be decreased by a factor of $\sim 1/2$. (We shall
see from Bethe's work below that it should actually be more like $\sim 1/3$.)
Bethe first made a fairly rough calculation of the four-body clusters, but
good enough to show when summed to all orders in $G$, it was tremendous,
giving
\begin{equation}
\Delta E_4\simeq -35 \mbox{ MeV/particle}
\end{equation}
to the binding energy.

Visiting CERN in the summer of 1964, Bethe learned of Fadeev's solution of the
three-body problem \cite{fad}. Hans found that he could write expressions for
the three-body problem, using the $G$-matrix as effective interaction; e.g.,
\begin{equation}
T\Phi=V\psi,
\end{equation}
with $T$ satisfying the equation
\begin{equation}
T=V+V\frac{1}{e}T
\end{equation}
where $e$ now represents the energy denominator of all three particles. The
$T$ can be split into
\begin{equation}
T=T^{(1)}+T^{(2)}+T^{(3)}
\end{equation}
but now one defines
\begin{equation}
T^{(1)}=V_{23}+V_{23}\frac{1}{e}T,
\end{equation}
etc. In other words, $T^{(1)}$ denotes that part of $T$ in which particle
``1'' did not take part in the last interaction. The ground state energy is
given by 
\begin{equation}
E_0=\langle \Phi \left| T \right| \Phi \rangle,
\end{equation}
where $\Phi$ is the unperturbed (free-particle) wave function. In short, one
could use the Brueckner $G$-matrices as effective interactions in the Fadeev
formalism, which then summed the three-body cluster terms to all orders in the
solution of these equations.

\begin{figure}
\begin{center}
\includegraphics[height=1in]{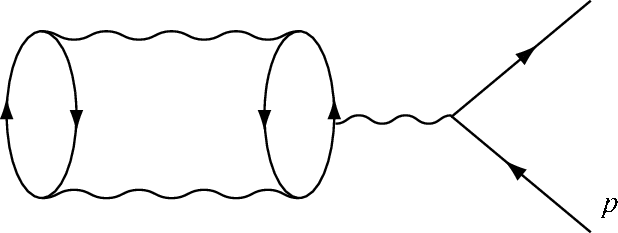}
\caption{The contribution of the three-body cluster to the off-shell self
  energy of a particle in state $p$. The wiggly line here represents the
  $G$-matrix. One particular contribution to this is shown in Fig.\
  \ref{offshell}.} 
\label{threebod}
\end{center}
\end{figure}

Hans came to Copenhagen in late summer of 1964, with the intention of solving
the Fadeev equations for the three-body system using the reference-spectrum
approximation eq.\ (\ref{rssps}) for the effective two-body interaction. David
Thouless, Hans' Ph.D. student in Cambridge, was visiting Nordita (Nordic
Institute of Theoretical Atomic Physics) at that time. I (G.E.B.) got David
together with Hans the morning after Hans arrived.\footnote{Our most fruitful 
  discussions invariably came the morning after he arrived in Copenhagen.}
Hans wrote down the three coupled equations on the blackboard and began to
solve them by some methods Dougy Rajaraman had used for summing four-body
clusters. David took a look at the three coupled equations with Hans'
$G$-matrix reference-spectrum approximation for the two-body interaction and
said ``These are just three coupled linear equations. Why don't you solve them
analytically?'' By late morning Hans had the solution, given in his 1965
paper. The result, which can be read from this paper, is that the off-shell
correction to the three-body cluster of Fig.\ \ref{threebod} should be cut
down by a 
factor of $1/3$. Hans' simple conclusion was ``when three nucleons are close
together, an elementary treatment would give us three repulsive pair
interactions. In reality, we cannot do more than exclude the wave function
from the repulsive region, hence we get only one core interaction rather than
3.''

In the summer of 1967, I (G.E.B.) gave the summary talk of the International
Nuclear Physics meeting held in Tokyo, Japan. This talk is in the
proceedings. I designed my comments as letters to the speakers, as did Herzog
in Saul Bellow's novel by that name.

\vspace{2pc}

{\it
\noindent ``Dear Professor Bethe,

\vspace{1pc}

\noindent First of all, your note is too short to be intelligible. But by
valiant efforts, and a high degree of optimism in putting together corrections
of the right sign, you manage to get within 3 MeV/particle of the binding
energy of nuclear-matter. It is nice that there are still some discrepancies,
because we must have some occupation for theorists, in calculating three-body
forces and other effects.

\vspace{1pc}

\noindent Most significantly, you confirm that it is a good approximation to
use plane-wave intermediate states in calculation of the G-matrix. This
simplifies life in finite nuclei immensely.

\vspace{1pc}

\noindent I cannot agree with you that there is no difference between hard- and
soft-core potentials. I remember your talk at Paris, where you showed that the
so-called dispersion term (the contribution of $G_s$ --G.E.B.), which is a
manifestation of off-energy-shell effects, differs by $\sim 3$ or $\sim 4$ MeV
for hard- and soft-core potentials, and that this should be the only
difference. I remain, therefore, a strong advocate of soft-core potentials,
and am confident that careful calculation will show them to be significantly
better.

\vspace{1pc}

\noindent Let me remind you that we (Kuo and Brown) always left out this
dispersion correction in our matrix elements for finite nuclei, in the hopes
of softer ones.

\vspace{1pc}

\noindent \dots \dots

\vspace{1pc}

\noindent Yours, etc."}

\vspace{2pc}

In summary, the off-energy-shell effects at the time of the reference-spectrum
through the dispersion correction $G_s$ contributed about $+6$ MeV to the
binding energy, a sizable fraction of the $\sim 15$ MeV total binding
energy. However, Bethe's solution of the three-body problem via Fadeev cut
this by $1/3$. The reference-spectrum paper was still using a vertical hard
core, whereas the introduction of the Yukawa-type repulsion by Breit
\cite{breit} and independently by Gupta \cite{gup}, although still 
involving Fourier (momentum) components of $p\sim m_\omega c$ with $m_\omega$
the 782 MeV/$c^2$ mass of the $\omega$-meson, $\sim 2\frac{1}{2}$ times
greater than $k_F$, 
cut the dispersion correction down somewhat more, so we were talking in 1967
about a remaining $\sim 1$ MeV compared with the $-16$ MeV binding energy per
particle. At that stage we agreed to neglect interactions in the particle
intermediate energy states, as is done in the Schwinger-Dyson Interaction
Representation. However, in the latter, any interactions that particles have
in intermediate states are expressed in terms of higher-order corrections,
whereas we just decided (``decreed'') in 1967 that these were negligible. An
extensive discussion of all of the corrections to the dispersion term from 
three-body clusters and other effects is given in Michael Kirson's Cornell
1966 thesis, written under Hans' direction. In detail, Kirson was not quite as
optimistic about dropping all off-shell effects as we have been above, but he
does find them to be at most a few MeV.

Once it was understood that the short-range repulsion came from the vector
meson exchange potentials, rather than a vertical hard core, the short-range
$G$-matrix $G_s$ in eq.\ (\ref{msg}) could be evaluated and it was found to
be substantially smaller than that for the vertical hard core. Because of the
smoothness in the potential, the off-shell effects were smaller, sufficiently
so that $G_s$ could be neglected. Thus, the short-range repulsion was
completely tamed and one had to deal with only the well-behaved power
expansion
\begin{equation}
G=v_l+v_l\frac{Q}{e}v_l+\cdots
\label{modg}
\end{equation}
In fact, in the $^1S_0$ states the first term gave almost all of the
attraction, whereas the iteration of the tensor force in the second term
basically gave the amount that the effective $^3S_1$ potential exceeded the
$^1S_0$ one in magnitude. (The tensor force contributed to only the triplet
states because it required $S=1$.)

Now Bethe's theory of the effective range in nuclear scattering \cite{bethe2}
showed how the scattering length and effective range, the first two terms in
the expansion of the scattering amplitude with energy, could be obtained from
any potential. The scattering length and effective range could be directly
obtained from the experimental measurements of the scattering.

Thus, one knew that any acceptable potential must reproduce these two
constants, and furthermore, from our previous discussion, contain a strong
short-range repulsion. To satisfy these criteria, Kallio and Kolltveit
\cite{kalkoll} therefore took singlet and triplet potentials
\begin{equation}
V_i(r) = \left \{ \begin{array}{cc}
        \infty & \mbox{for $r \leq 0.4$ fm} \\
        -A_ie^{-\alpha_i(r-0.4)} & \mbox{for $r > 0.4$ fm} \\
        \end{array} \right.  \mbox{ for $i = s,t$,}
\label{kkpot}
\end{equation}
where
\begin{equation}
\begin{array}{cc}
        A_s = 330.8 \mbox{ MeV}, & \alpha_s = 2.4021 \mbox{ $\rm fm^{-1}$} \\
        A_t = 475.0 \mbox{ MeV}, & \alpha_t = 2.5214 \mbox{ $\rm fm^{-1}$} \\
\end{array}
\end{equation}
By using this potential, Kallio and Kolltveit obtained good fits to the
spectra of light nuclei.

Surprising developments, summarized in the {\it Physics Report}
``Model-Independent Low Momentum Nucleon Interaction from Phase Shift
Equivalence'' \cite{bogner}, showed how one could define an effective
interaction between nucleons by starting from a renormalization group approach
in which momenta larger than a cutoff $\Lambda$ are integrated out of the
effective $G$-matrix interaction. In fact, the data which went into the phase
shifts obtained by various groups came from experiments with center of mass
energies less than 350 MeV, which corresponds to a particle momentum of 2.1
fm$^{-1}$. Thus, setting $\Lambda$ equal to 2.1 fm$^{-1}$ gave an effective
interaction $V_{\rm low-{\it k}}$ which included all experimental measurements.

\begin{figure}
\begin{center}
\includegraphics[height=2.7in]{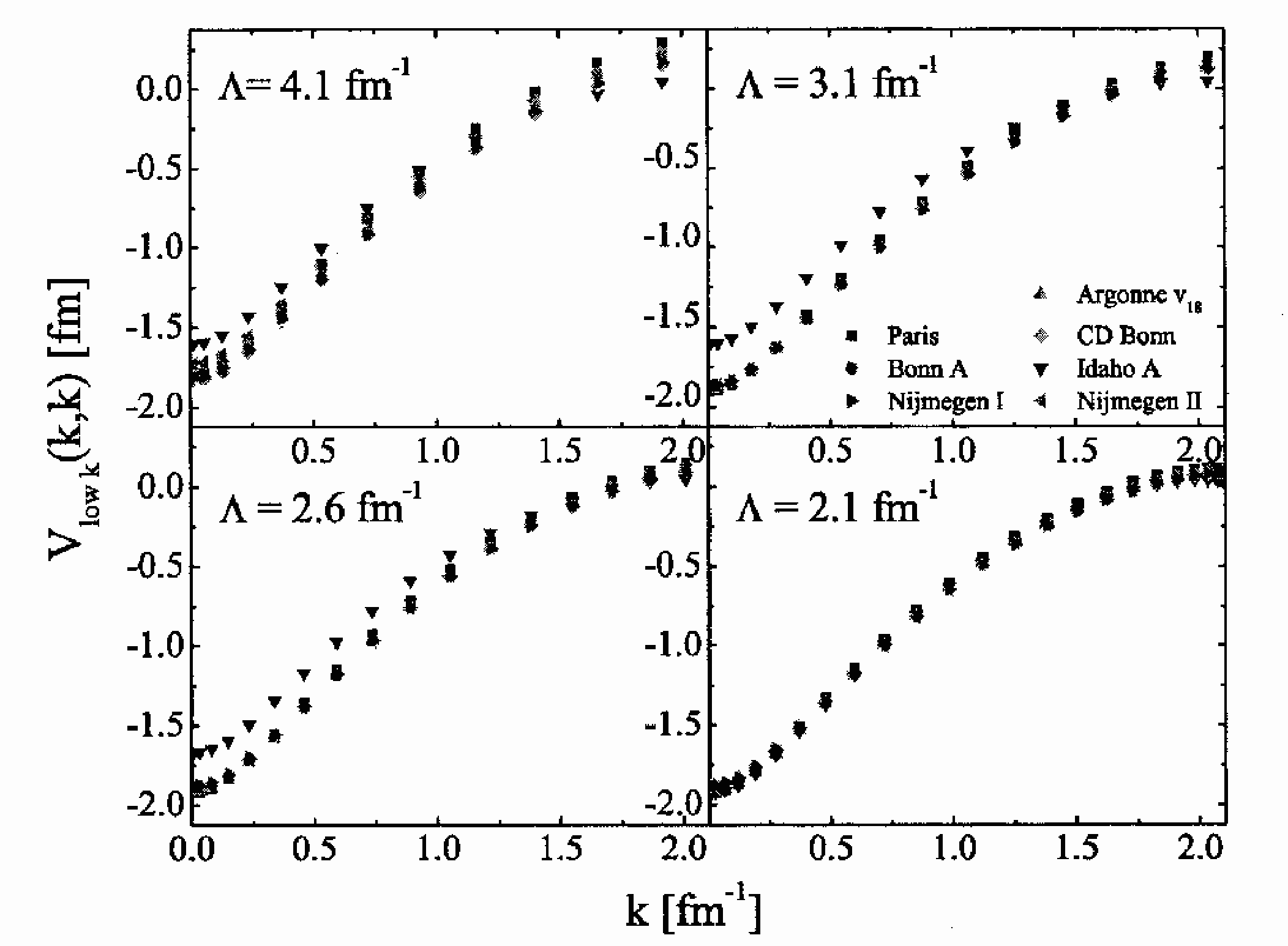}
\caption{Diagonal matrix elements of $V_{\rm low-{\it k}}$ for different
  high-precision potentials in the $^1S_0$
  partial wave with various cutoffs $\Lambda$.}
\label{vlowk1}
\end{center}
\end{figure}

\begin{figure}
\begin{center}
\includegraphics[height=2.7in]{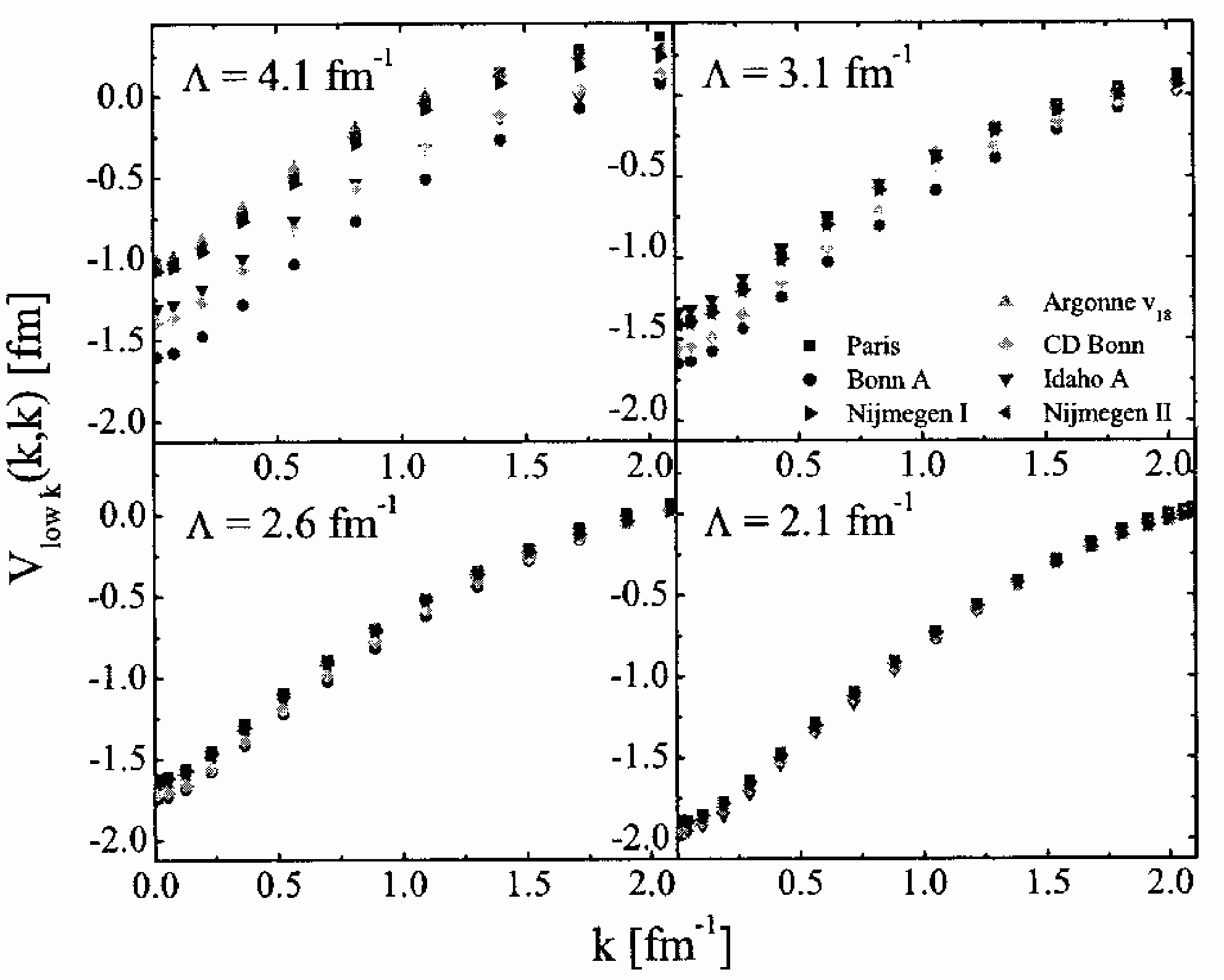}
\caption{Diagonal matrix elements of $V_{\rm low-{\it k}}$ for different
  high-precision potentials in the $^3S_1$
  partial wave with various cutoffs $\Lambda$.} 
\label{vlowk2}
\end{center}
\end{figure}

We show in Figs.\ \ref{vlowk1} and \ref{vlowk2} (Figs.\ 11 and 12 in
\cite{bogner}) the collapse of the 
various potentials obtained from different groups as $\Lambda$ is lowered from
4.1 fm$^{-1}$ to 2.1 fm$^{-1}$. The reason for this collapse is that as the
data included in the phase shift determinations is lowered from a large energy
range to that range which includes only measured data; the various groups were
all using the same data in order to determine their scattering amplitudes and
therefore obtained the same results. Because of this uniqueness, $V_{\rm
  low-{\it k}}$ is much used in nuclear structure physics these days.


\begin{figure}[ht]
\begin{minipage}{18pc}
\includegraphics[width=18pc]{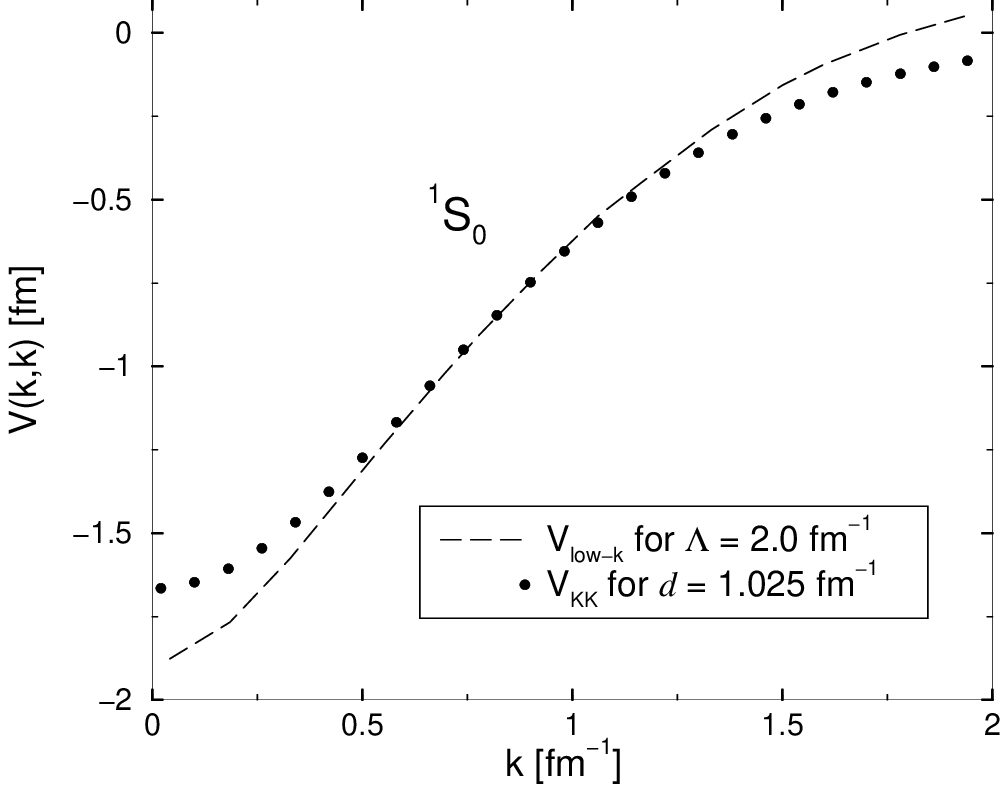}
\caption{The $^1S_0$ diagonal matrix elements of $V_{\rm low-{\it k}}$ and the
  Kallio-Kolltveit potential for a configuration space cutoff of 1.025 fm.}
\label{singlet}
\end{minipage}\hspace{2pc}
\begin{minipage}{18pc}
\includegraphics[width=18pc]{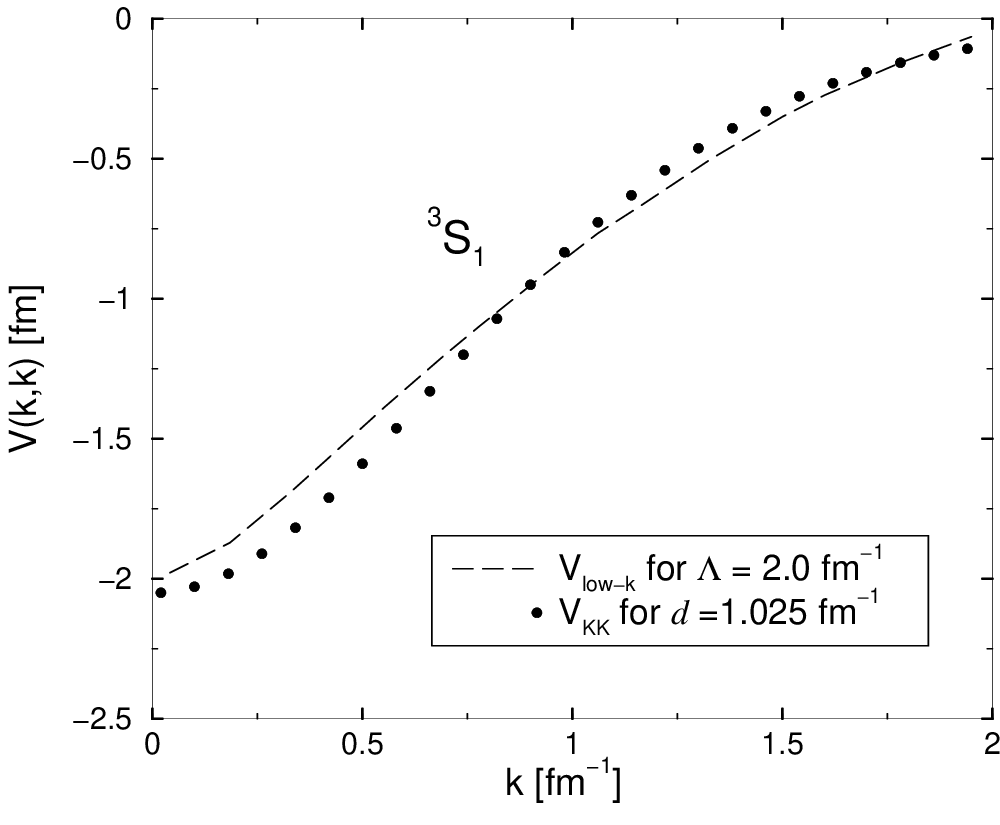}
\caption{The $^3S_1$ diagonal matrix elements of $V_{\rm low-{\it k}}$ and the
  Kallio-Kolltveit potential for a configuration space cutoff of 1.025 fm.} 
\label{triplet}
\end{minipage}
\end{figure}

Now the above renormalization group calculations were all carried out in
momentum space. What is the relation of $V_{\rm low-{\it k}}$ to the effective
interaction of eq.\ (\ref{modg})? Although in the Mozskowski-Scott method the
separation distance $d$ should be obtained separately for each channel, for
comparison with $V_{\rm low-{\it k}}$ (which has the same momentum cutoff in
all channels) it is most convenient to 
determine it for the $^1S_0$ channel ($d = 1.025$) and then use the same $d$
in the $^3S_1$ channel. We show a comparison between $V_{\rm low-{\it k}}$ and
the $G$ of eq.\ (\ref{modg}) in Figs.\ \ref{singlet} and \ref{triplet}.

\begin{figure}[thb]
\begin{center}
\includegraphics[height=2.5in]{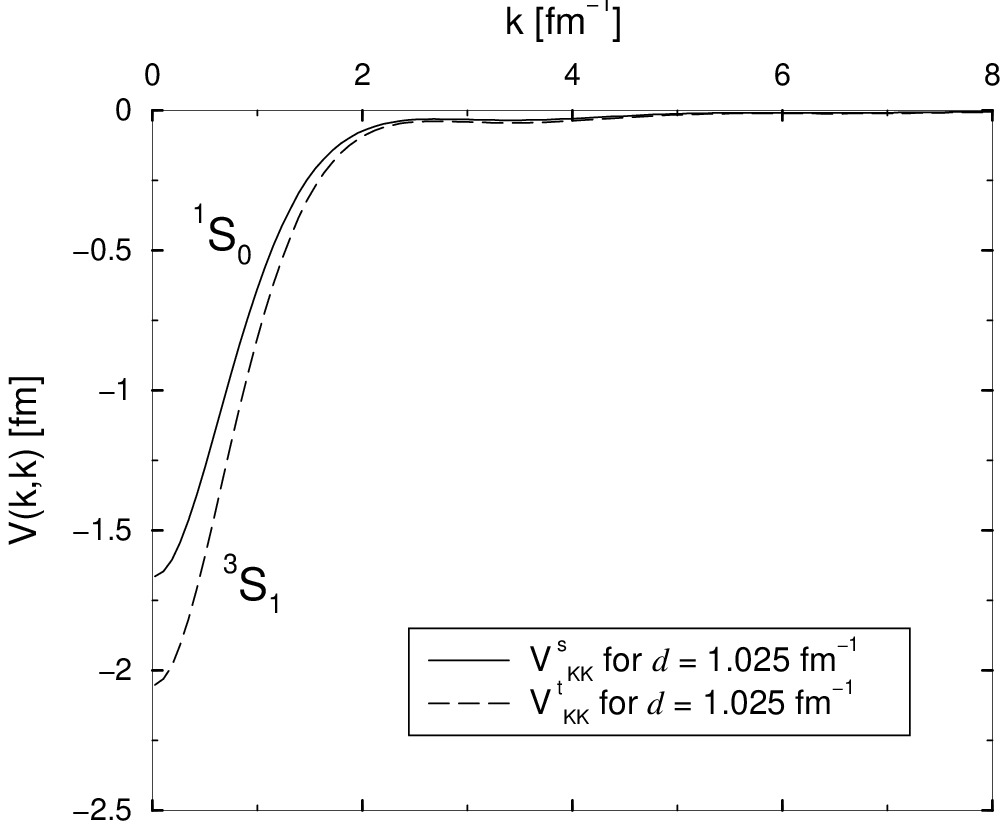}
\caption{Diagonal matrix elements of the long-distance Kallio-Kolltveit
  potential $v_l$, including momenta above the $V_{\rm low-{\it k}}$ cutoff of
  2.1 fm$^{-1}$.} 
\label{vhighk}
\end{center}
\end{figure}

In Fig.\ \ref{vhighk} we show the diagonal matrix elements of the
Kallio-Kolltveit 
potential, essentially the $G$ of eq.\ (\ref{modg}). Note that although the
separation of short and long distance was made in configuration space, there
are essentially no Fourier components above $k\sim 2 \mbox{ fm}^{-1}$ in
momentum space.

It is, therefore, clear that the $G$ of eq.\ (\ref{modg}) that Bethe ended
with is essentially $V_{\rm low-{\it k}}$, the only difference being that he
made the separation of scales in configuration space whereas in $V_{\rm
  low-{\it k}}$ it is made in momentum space. Thus, we believe that Hans Bethe
arrived at the right answer in ``The Nuclear Many Body Problem,'' but only
much later did research workers use it to fit spectra.

\end{document}